\pdfoutput=1

\documentclass[conference]{IEEEtran}
\usepackage{fancyhdr}
\ifCLASSINFOpdf
\usepackage[pdftex]{graphicx}
\graphicspath{{./}}
\DeclareGraphicsExtensions{.pdf,.jpeg,.png}
\else
\fi
\usepackage[cmex10]{amsmath}
\usepackage[caption=false,font=footnotesize]{subfig}

\usepackage{url}

\begin{document}
%
\title{Parallelizing the QUDA Library for Multi-GPU Calculations in Lattice Quantum Chromodynamics}

\author{\IEEEauthorblockN{Ronald Babich}
\IEEEauthorblockA{Center for Computational Science\\
Boston University\\
Boston, Massachusetts 02215, USA\\
Email: rbabich@bu.edu}
\and
\IEEEauthorblockN{Michael A. Clark}
\IEEEauthorblockA{Harvard-Smithsonian Center for Astrophysics\\
Cambridge, Massachusetts 02138, USA\\
Email: mikec@seas.harvard.edu}
\and
\IEEEauthorblockN{B\'alint Jo\'o}
\IEEEauthorblockA{Jefferson Laboratory\\
Newport News, Virginia 23606, USA\\
Email: bjoo@jlab.org}}


%


\maketitle

\thispagestyle{fancy}
\lhead{}
\rhead{}
\chead{}
\lfoot{\footnotesize{\copyright 2010 IEEE Personal use of
this material is permitted. However, permission to
reprint/republish
this
material
for
advertising
or
promotional purposes or for creating new collective works
for resale or redistribution to servers or lists, or to
reuse any copyrighted component of this work in other
works must be obtained from the IEEE. \newline
SC10 November 2010, New Orleans, Louisiana, USA 978-1-
4244-7558-2/10/\$26.00}}
\rfoot{}
\cfoot{}
\renewcommand{\headrulewidth}{0pt}
\renewcommand{\footrulewidth}{0pt}

\begin{abstract}
  Graphics Processing Units (GPUs) are having a transformational
  effect on numerical lattice quantum chromodynamics (LQCD)
  calculations of importance in nuclear and particle physics.  The
  QUDA library provides a package of mixed precision sparse matrix
  linear solvers for LQCD applications, supporting single GPUs based
  on NVIDIA's Compute Unified Device Architecture (CUDA).  This
  library, interfaced to the QDP++/Chroma framework for LQCD
  calculations, is currently in production use on the ``9g'' cluster at
  the Jefferson Laboratory, enabling unprecedented price/performance
  for a range of problems in LQCD.  Nevertheless, memory constraints
  on current GPU devices limit the problem sizes that can be tackled.
  In this contribution we describe the parallelization of the QUDA
  library onto multiple GPUs using MPI, including strategies for the
  overlapping of communication and computation.  We report on both
  weak and strong scaling for up to 32 GPUs interconnected by
  InfiniBand, on which we sustain in excess of 4 Tflops.
\end{abstract}


%
\IEEEpeerreviewmaketitle

\section{Introduction}

Lattice quantum chromodynamics (LQCD) is the lattice discretized
theory of the strong nuclear force, the force that binds
quarks together into particles such as the proton and neutron.  High precision
predictions from LQCD are required for testing the standard model of
particle physics, a task with increased importance in the era of the
Large Hadron Collider (LHC), where deviations between numerical LQCD
predictions and experiment could be signs of new physics. LQCD also
has a vital role to play in nuclear physics, where such calculations
are used to compute and classify the excited states of protons,
neutrons and other hadrons; to study hadronic structure; and to
compute the forces and binding energies in light nuclei.

LQCD is a grand challenge subject, with large-scale computations
consuming a considerable fraction of publicly available supercomputing
resources. The computations typically proceed in two phases: in the
first phase, one generates thousands of {\em configurations} of the
strong force fields (gluons), colloquially referred to as {\it gauge
  fields}.  This computation is a long-chain Monte Carlo process,
requiring the focused power of leadership class computing facilities
for extended periods. In the second phase, these configurations are
{\em analyzed}, a process that involves probing the interaction of
quarks and gluons with each other on each configuration. The
interactions are calculated by solving systems of linear equations
with coefficients determined by elements of the gauge field.  On each
configuration the equations are solved for many right hand sides, and
the solution vectors are used to compute the final observables of
interest. This second phase can proceed independently on each
configuration, and as a result, cluster partitions of modest size have
proven to be highly cost-effective for this purpose. Until a few years
ago, the analysis phase would often account for a relatively small
part of the cost of the overall calculation, with analysis
corresponding to perhaps 10\% of the cost of gauge field generation.
In recent years, however, focus has turned to more challenging
physical observables and new analysis techniques that demand solutions
to the aforementioned linear equations for much larger numbers of
right hand sides (see, e.g.,~\cite{Babich:2009rq,Peardon:2009gh}).  As
a result, the relative costs have shifted to the point where analysis
often requires an equal or greater amount of computation than gauge
field generation.

The rapid growth of floating point power in graphics processing units
(GPUs) together with drastically improved tools and programmability
has made GPUs a very attractive platform for LQCD computations.  The
QUDA library~\cite{quda-webpage,Clark:2009wm} provides a package of
optimized kernels for LQCD that take advantage of NVIDIA's Compute
Unified Device Architecture (CUDA).  Once coupled to LQCD application
software, e.g., Chroma~\cite{Edwards:2004sx}, this provides a powerful
framework for lattice field theorists to exploit.  The ``9g'' GPU
cluster at Jefferson Laboratory features 192 NVIDIA GTX 285 GPUs
providing over 30 Tflops of sustained performance in LQCD, when
aggregated over single GPU jobs.  For problems that can be accommodated
by the limited GPU memory, the price/performance compared to typical
clusters or massively parallel supercomputers (e.g., BlueGene/P) is
improved by around a factor of five.  However, for problem sizes that
are too large, individual GPUs have no benefit.

Even for problems that do fit on a single GPU, the economics of
constructing a GPU cluster tend to motivate provisioning each cluster
node with multiple GPUs, since the incremental cost of an additional
GPU is fairly small.  In this scenario, it is possible to run multiple
independent jobs on each node, but then the size of the host memory
may prove to be the limiting constraint.  The obvious recourse in both
cases is therefore to parallelize a single problem over multiple GPUs,
which is the subject of our present work.

The paper is organized as follows.  In Sections~\ref{sec:LQCD}
and~\ref{sec:GPUs} we review basic details of the LQCD application and
of NVIDIA GPU hardware.  We then briefly consider some related work in
Section~\ref{sec:related-work} before turning to a general description
of the QUDA library in Section~\ref{sec:QUDA}. Our parallelization of
the quark interaction matrix is described in \ref{sec:multi-gpu}, and we
present and discuss our performance data for the parallelized solver
in Section~\ref{sec:solver-perf}. We finish with conclusions and a
discussion of future work in Section~\ref{sec:conclusions}.

\section{Lattice QCD}\label{sec:LQCD}

The necessity for a lattice discretized formulation of QCD arises due
to the failure of perturbative approaches commonly used for
calculations in other quantum field theories, such as electrodynamics.
Quarks, the fundamental particles that are at the heart of QCD, are
described by the Dirac operator acting in the presence of a local
SU(3) symmetry.  On the lattice, the Dirac operator becomes a large
sparse matrix, \(M\), and the calculation of quark physics is
essentially reduced to many solutions to systems of linear equations
given by
\begin{equation}
Mx = b.
\label{eq:linear}
\end{equation}
The form of $M$ on which we focus in this work is the
Sheikholeslami-Wohlert \cite{Sheikholeslami:1985ij} (colloquially
known as {\em Wilson-clover}) form, which is a central difference
discretization of the Dirac operator.  When acting in a
vector space that is the tensor product of a 4-dimensional discretized
Euclidean spacetime, {\it spin} space, and {\it color} space it is
given by
\begin{align}
 M_{x,x'} &= - \frac{1}{2} \displaystyle \sum_{\mu=1}^{4} \bigl(
 P^{-\mu} \otimes U_x^\mu\, \delta_{x+\hat\mu,x'}\, + P^{+\mu} \otimes
 U_{x-\hat\mu}^{\mu \dagger}\, \delta_{x-\hat\mu,x'}\bigr) \nonumber \\
&\quad\, + (4 + m + A_x)\delta_{x,x'} \nonumber \\
&\equiv  - \frac{1}{2}D_{x,x'} + (4 + m + A_{x}) \delta_{x,x'}.
\label{eq:M}
\end{align} 
Here \(\delta_{x,y}\) is the Kronecker delta; \(P^{\pm\mu}\) are
\(4\times 4\) matrix projectors in {\it spin} space; \(U\) is the QCD
gauge field which is a field of special unitary $3\times 3$ (i.e.,
SU(3)) matrices acting in {\it color} space that live between the
spacetime sites (and hence are referred to as link matrices); \(A_x\)
is the \(12\times12\) clover matrix field acting in both spin and
color space,\footnote{Each clover matrix has a Hermitian block
  diagonal, anti-Hermitian block off-diagonal structure, and can be
  fully described by 72 real numbers.} corresponding to a first order
discretization correction; and \(m\) is the quark mass parameter.  The
indices \(x\) and \(x'\) are spacetime indices (the spin and color
indices have been suppressed for brevity).  This matrix acts on a
vector consisting of a complex-valued 12-component \emph{color-spinor}
(or just {\em spinor}) for each point in spacetime.  We refer to the
complete lattice vector as a spinor field.

\begin{figure}[h]
\begin{center}
\includegraphics[width=2.5in]{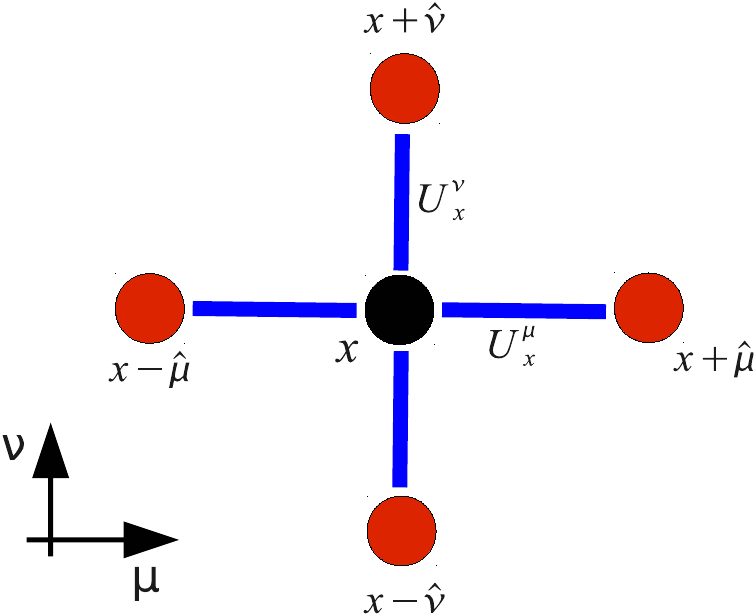}
\end{center}
\caption{\label{fig:dslash}The nearest neighbor stencil part of the
  lattice Dirac operator $D$, as defined in (\ref{eq:M}), in the $\mu-\nu$
  plane.  The \emph{color-spinor} fields are located on the
  sites. The SU(3) color matrices $U^\mu_x$ are associated with the links. The
  nearest neighbor nature of the stencil suggests a natural even-odd (red-black)
  coloring for the sites.}
\end{figure}

Since \(M\) is a large sparse matrix, an iterative Krylov solver is
typically used to obtain solutions to (\ref{eq:linear}), requiring
many repeated evaluations of the sparse matrix-vector product.  The
matrix is non-Hermitian, so either Conjugate Gradients
\cite{Hestenes:1952} on the normal equations (CGNE or CGNR) is used,
or more commonly, the system is solved directly using a non-symmetric
method, e.g., BiCGstab \cite{vanDerVorst:1992}.  Even-odd (also known
as red-black) preconditioning is used to accelerate the solution
finding process, where the nearest neighbor property of the
\(D_{x,x'}\) matrix (see Fig.~\ref{fig:dslash}) is exploited to solve
the Schur complement system~\cite{DeGrand:1990}.  This has no effect
on the overall efficiency since the fields are reordered such that all
components of a given parity are contiguous.  The quark mass controls
the condition number of the matrix, and hence the convergence of such
iterative solvers.  Unfortunately, physical quark masses correspond to
nearly indefinite matrices.  Given that current leading lattice
volumes are \(32^3\times256\), for \(>10^8\) degrees of freedom in
total, this represents an extremely computationally demanding task.



\section{Graphics Processing Units} \label{sec:GPUs}

In the context of general-purpose computing, a GPU is effectively an
independent parallel processor with its own locally-attached memory,
herein referred to as \emph{device memory}.  The GPU relies on the
host, however, to schedule blocks of code (or \emph{kernels}) for
execution, as well as for I/O.  Data is exchanged between the GPU and
the host via explicit memory copies, which take place over the
PCI-Express bus.  The low-level details of the data transfers, as well
as management of the execution environment, are handled by the GPU
device driver and the runtime system.

It follows that a GPU cluster embodies an inherently heterogeneous
architecture.  Each node consists of one or more processors (the CPU)
that is optimized for serial or moderately parallel code and attached
to a relatively large amount of memory capable of tens of GB/s of
sustained bandwidth.  At the same time, each node incorporates one or
more processors (the GPU) optimized for highly parallel code attached
to a relatively small amount of very fast memory, capable of 150 GB/s
or more of sustained bandwidth.  The challenge we face is that these
two powerful subsystems are connected by a narrow communications
channel, the PCI-E bus, which sustains at most 6 GB/s and often less.
As a consequence, it is critical to avoid unnecessary transfers
between the GPU and the host.  For single-GPU code, the natural
solution is to carry out all needed operations on the GPU; in the QUDA
library, for example, the linear solvers are written such that the
only transfers needed are the initial upload of the source vector to
the GPU and the final download of the solution, aside from occasional
small messages needed to complete global sums.  A multi-GPU
implementation, however, cannot avoid frequent large data transfers, and so the
challenge becomes to overlap the needed communication with useful
work.  This is exacerbated further if one wishes to take advantage of
many GPUs spread across multiple nodes, since the bandwidth provided
the fastest available interconnect, QDR InfiniBand, is half again that
provided by (x16) PCI-E.

\begin{table}
\begin{tabular}{|l|c|c|c|c|c|} \hline
 &  &  GB/s & \multicolumn{2}{|c|}{Gflops} &  GiB \\ \hline 
Card & Cores & \hspace{-2mm} Bandwidth \hspace{-2mm} & 32-bit & 64-bit & RAM  \\ \hline
GeForce 8800 GTX & 128 & 86.4 &  518 & N/A & 0.75 \\ \hline
Tesla C870       & 128 & 76.8 &  518 & N/A & 1.5 \\ \hline
GeForce GTX 285  & 240 & 159  & 1062 &  88 & 1.0 - 2.0 \\ \hline
Tesla C1060      & 240 & 102  &  933 &  78 & 4.0 \\ \hline
GeForce GTX 480  & 480 & 177  & 1345 & 168 & 1.5 \\ \hline
Tesla C2050      & 448 & 144  & 1030 & 515 & 3.0 \\ \hline
\end{tabular}
\caption{\label{table:specs}Specifications of representative NVIDIA
  graphics cards.}
\end{table}

We turn now to the architecture of the GPU itself.  Our purpose is
only to highlight those features that have directly influenced our
implementation.  We focus here on cards produced by NVIDIA and
specifically on the GT200 generation, as typified by the Tesla C1060
and the GeForce GTX 285, since the latter will serve as our test bed.
The GT200 series is the second of the three extant generations of
CUDA-enabled cards, representative examples of which are listed in
Table~\ref{table:specs}.  The most recent generation, embodying
NVIDIA's ``Fermi'' architecture, is only now becoming available in
mid-2010.  We note that while hardware features and performance differ
between generations, these have relatively little impact on our
multi-GPU strategy.  Likewise, most of the considerations we discuss
would apply even to an OpenCL implementation targeting graphics
cards produced by AMD/ATI.

GPUs support a single-program multiple-data (SPMD) programming model
with up to thousands of threads in flight at once.  Each thread
executes the same kernel, using a unique thread index to determine the
work that should be carried out.  The GPU in the GeForce GTX 285 card
consists of 240 cores organized into 30 multiprocessors of 8 cores
each.  Each core services multiple threads concurrently by alternating
between them on successive clock cycles, so a group of 32 threads (a
{\em warp} in NVIDIA's parlance) is executing on the multiprocessor at
a given moment.  At the same time, many additional threads (ideally
hundreds) are typically resident on the multiprocessor and ready to
execute.  This allows the multiprocessor to swap in a new set of 32
threads when a given set stalls -- while waiting for a memory access
to complete, for example.  In order to hide latency, it is desirable
to have many threads resident at once, but each such thread requires a
certain number of registers and quantity of shared memory, which
limits the total.  Just as on a CPU, a \emph{register} is where a
variable is stored while it is being operated on or written out.
Registers are not shared between threads.  \emph{Shared memory}, on
the other hand, may be shared between threads executing on the same
multiprocessor.  Strictly speaking, the threads must belong to the
same \emph{thread block}, a group of threads whose size is specified
by the programmer; each thread block must consist of a multiple of 64
threads, and one or more thread blocks may be active on a
multiprocessor at a time.  The GeForce GTX 285 provides 16,384
single-precision registers (8,192 in double precision) and 16~KiB of
shared memory per multiprocessor.

The CUDA programming model treats the threads within a block as
independent threads of execution, as though they were executing on
cores that were true scalar processors; threads may take independent
code paths, read arbitrary locations in memory, and so on.  In order
to obtain optimal performance, however, it is better to treat the
multiprocessor as a single 32-lane or 16-lane SIMD unit.  This follows
from two considerations.  First, when threads within a set of 32 (a
warp) take different paths at a branch, the various paths are
serialized and executed one after another, a condition known as ``warp
divergence.''  Second, when accessing device memory, maximum bandwidth
is achieved only when 16 threads access contiguous elements of memory,
where each such element is a 4-byte, 8-byte, or 16-byte block.  (The
CUDA C language defines various short vector types for this purpose,
e.g., {\em float2, float4, double2, short4,} etc.)  This allows the
transfer to proceed as a single ``coalesced'' memory transaction.  As
described in Section~\ref{sec:QUDA} below, this consideration directly
influences the layout of our data.

An additional consideration has to do with the physical organization
of the device memory.  Like many classic vector architectures but
unlike commodity CPUs, GPUs are equipped with a very wide memory bus
(512-bit on the GTX 285) with memory partitioned into multiple banks
(eight on the GTX 285).  Successive 256-byte regions in device memory
map to these partitions in a round-robin fashion.  This organization
is generally transparent to the programmer, but if memory is accessed
with a stride that results in traffic to only a subset of the
partitions, performance will be lower than if all partitions were
stressed equally.  Such ``partition camping'' can result in an
unexpected loss of performance for certain problem
sizes~\cite{paulius}.  As discussed in~\cite{Clark:2009wm} and
Section~\ref{sec:QUDA} below, this was found to be a problem for
certain lattice volumes in our LQCD application, with the solution
being to pad the relevant arrays to avoid the camping.

To summarize, the GPU memory hierarchy consists of globally-accessible
device memory and local per-multiprocessor shared memory, often used
as a manually-managed cache, as well as local registers.  In
addition, GPUs such as the GTX 285 provide two special-purpose caches.
The first is a read-only texture cache, which speeds up reads from
device memory for certain kinds of access patterns.  It also provides
various addressing modes and rescaling capability.  As described
further in Section~\ref{sec:QUDA}, we take advantage of the latter in
our half-precision implementation.  Finally, each multiprocessor
provides a small \emph{constant cache} (8~KiB on the GTX 285), which
is useful for storing run-time parameters and other constants,
accessible to all threads with very low latency.

\section{Related Work}\label{sec:related-work}

GPUs were first used to perform LQCD calculations in
\cite{Egri:2006zm}. This pioneering study predated various
programmability improvements, such as the C for CUDA framework, and
hence was implemented using the OpenGL graphics API.  It targeted
single GPU devices. The QUDA library~\cite{quda-webpage} was discussed
extensively in \cite{Clark:2009wm,Barros:2008rd}, where the primary
techniques and algorithms for maximizing the efficient use of memory
bandwidth were presented for a single GPU device. LQCD on GPUs has
also been explored in \cite{Ibrahim-2008}, which focused on questions
of fine grained vs.\ coarse grained parallelism on single GPU
devices. In addition, there are several as yet unpublished efforts
aimed at exploiting GPUs for LQCD underway.

LQCD has also been implemented on other heterogeneous devices,
primarily on the Cell Broadband Engine. Efforts in this direction have
been reported in \cite{Belletti:2007pp,Baier:2009yq} as part of the
``QCD Parallel Computing on the Cell Broadband Engine'' (QPACE) project
and elsewhere~\cite{Shi:2009uq,Spray:2008nt}.

Outside the context of LQCD, general challenges of implementing
message passing on heterogeneous architectures have been considered
for GPUs in \cite{Stuart:2009:MPO} and for the RoadRunner
supercomputer in \cite{Parkin2009:rev-accel}. An effort to provide a
general message passing framework utilizing CUDA, MPI, and POSIX
threads is also underway at Jefferson Laboratory~\cite{Chen:2010}.

\section{QUDA}\label{sec:QUDA}

The QUDA library is a publicly available collection of optimized QCD
kernels built on top of the CUDA framework, with a simple C interface
to allow for easy integration with LQCD application software.
Currently, QUDA provides highly optimized CG and BiCGstab linear
solvers for a variety of different discretizations of the Dirac
operator, as well as other time critical components.

\subsection{General Strategy}
The power of GPUs may only be brought to bear when a large degree of
parallelism is available.  LQCD is fortunate in this regard, since
parallelism can easily be achieved by assigning one thread to
each lattice site.  The mapping from the linear thread index to the
4-dimensional spacetime index is easily obtained through integer
division and modular arithmetic involving the lattice dimensions.
These runtime parameters (and others, such as boundary conditions) are
stored in the constant cache.

In applying the lattice Dirac operator, each thread is thus
responsible for gathering its eight neighboring spinors (24
numbers apiece), applying the appropriate spin projector for each,
multiplying by the color matrix connecting the sites (18
numbers), and accumulating the results together with the local spinor
(24 numbers) weighted by the mass term.  The Wilson-clover
discretization also requires an extra multiplication by the clover matrix
(72 numbers) before the result (24 numbers) is saved to memory.  In
total, the application of the Wilson-clover matrix requires 3696
floating point operations for every 2976 bytes of memory traffic in
single precision (assuming kernel fusion to minimize memory traffic).

\subsection{Field Ordering}
The ordering typical on a CPU is to place the spacetime dimensions
running slowest, with internal dimensions (color, spin, and real/imaginary)
running fastest.  However, since memory coalescing is only achieved if
adjacent threads load consecutive blocks of 4, 8, or 16 bytes, the
fields must be reordered to ensure this condition.  This can be
achieved if we abandon the naive ordering,
\begin{equation}
i_{old} = x N_{int} + n,
\end{equation}
in favor of the new mapping
\begin{equation}
i_{new} = N_{vec} \left( V \left\lfloor \frac{n}{N_{vec}} \right\rfloor
+ x \right) + n \,\mbox{mod}\, N_{vec}.
\end{equation}
Here $V$ is the spacetime volume; \(x\) is the linear spacetime index
running from 0 through $V-1$; \(n\) corresponds to the internal index
running from 0 through $N_{int}-1$, with \(N_{int}=\) 24, 12, and 72
elements for the spinor, color (see Section~\ref{sec:12gauge}), and
clover fields respectively; and \(N_{vec}\) is the length of the vector
type used (e.g., \(N_{vec}=1, 2, 4\) for {\em
  float, float2,} and {\em float4}).  We have found that using \(N_{vec}=4\)
and \(N_{vec}=2\) is optimal in single and double precision,
respectively, each corresponding to a length of 16 bytes.

\begin{figure}[h]
\begin{center}
\includegraphics[width=2.3in]{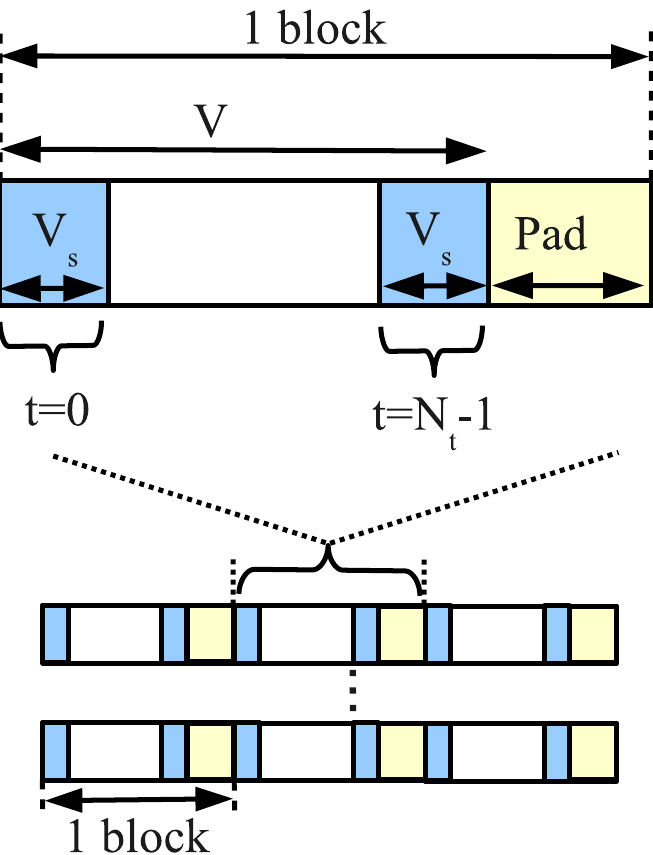}
\end{center}
\caption{\label{fig:layout}The field ordering used in QUDA: $N_{int}
  V$ numbers are broken up into $N_{int}/N_{vec}$ blocks of $V$ short
  vectors ( $=N_{vec} V$ numbers). Successive threads thus read
  successive short vectors ensuring coalescing of the memory
  transfers. Within a block the time index runs slowest, implying that
  the two faces on the temporal boundaries are each contiguous
  within the block; each face is stored in $V_s$ vectors. The blocks
  are separated by a padding region to avoid partition camping. As an
  example, in single precision one would use the {\em float4} vector
  type ($N_{vec}=4$), and thus 6 blocks would be needed to store the $24V$
  numbers that make up a color-spinor. Likewise, in 2-row storage, the
  gauge field would need 3 blocks to store the $12V$ numbers needed
  for each direction $\mu$. With 4 such directions, altogether 12
  blocks are needed to store all the link matrices. With the size of
  the padding chosen to be $V_s= XYZ$ sites, the ghost zone of link
  matrices $U^{\mu}_{x-\hat{\mu}}$ can be hidden entirely in the
  padding.}
\end{figure}

QUDA follows the usual lattice site assignment for the color matrices.
The color matrix connecting sites \(x\) and \(x+\hat{\mu}\) is denoted
by \(U_x^\mu\) and stored at lattice site \(x\).  It follows that
\(U_{x-\hat\mu}^{\mu}\), which is required for the gather from the
backwards direction for site \(x\), is stored at site \(x-\hat{\mu}\).
(The matrix conjugation is performed at no cost through register
relabeling in the kernel.)

As anticipated in Section~\ref{sec:GPUs}, for certain problem sizes
performance may be affected by partition camping.  The simple solution
QUDA takes to this problem is to pad the gauge, spinor, and clover
fields by one spatial volume, \(V_s = XYZ\), so that the linear indexing
is given by
\begin{equation}
i  = N_{vec} \left( (T+1)V_s \left\lfloor \frac{n}{N_{vec}} \right\rfloor
+ x \right) + n \,\mbox{mod}\, N_{vec}.
\end{equation}
Here \(X\), \(Y\), \(Z\) and the \(T\) are the lengths of the
respective spacetime dimensions, with \(V = V_sT\).  Although not
originally intended for this purpose, padding the fields by an extra
spatial volume is also convenient for the parallelization process (see
Section~\ref{sec:multi-gpu}).  We illustrate the field ordering in
Fig.~\ref{fig:layout}.

\subsection{Reducing Memory Traffic}
Given the peak instruction and bandwidth throughputs of current GPUs
(Table \ref{table:specs}), evaluation of the Wilson-clover matrix
vector product is strongly bandwidth bound.  The approach taken by
QUDA is to minimize memory traffic, even at the expense of additional
floating point operations, to accelerate performance using the
following techniques:
\subsubsection {Gauge field compression}\label{sec:12gauge} Only the first two rows of
the color matrices are stored in device memory, and using
unitarity, the third row is reconstructed in registers from the
complex conjugate of the cross product of the first two rows.
\subsubsection {Similarity transformations}\label{sec:similar}
Physically motivated similarity transformations are employed to increase
the sparsity of the matrix.  In particular, the spin projectors in the
temporal dimension \(P^{\pm4}\) are diagonalized by changing from the
conventional chiral basis to a ``non-relativistic'' basis,
\begin{equation}
P^{\pm 4} = 
\left(\begin{array}{rrrr}
  1&0&\pm1&0\\
  0&1&0&\pm1\\
  \pm1&0&1&0\\
  0&\pm1&0&1\\
\end{array}\right) \Longrightarrow
\end{equation}
\[
P^{+4} = 
\left(\begin{array}{rrrr}
  2&0&0&0\\
  0&2&0&0\\
  0&0&0&0\\
  0& 0&0&0\\
\end{array}\right), 
P^{-4} = 
\left(\begin{array}{rrrr}
  0&0&0&0\\
  0&0&0&0\\
  0&0&2&0\\
  0& 0&0&2\\
\end{array}\right).
\]
This has the benefit that only 12 real numbers need be loaded when
gathering neighboring spinors in the temporal direction and also aids
our parallelization approach (See Section~\ref{sec:multi-gpu}).
\subsubsection{Precision Truncation} Further acceleration is obtained
through the use of 16-bit fixed point storage, from here on referred
to as half precision.  This is implemented by reading the gauge field
and spinor field elements via the texture cache, using the read mode
{\it cudaReadModeNormalizedFloat}.  When a texture reference is
defined using this mode, a signed 16-bit (or even 8-bit) integer read
in from device memory will be automatically converted to a 32-bit
floating point number in the range \([-1,1]\).  This format is
immediately suitable for the color matrices since all of their
elements lie exactly in this range, as a consequence of unitarity.
The spinors require an extra normalization, which is shared between
all elements of a single spinor.\footnote{This sharing of a common
  normalization factor among spinor elements is motivated by the fact
  that the action of the Wilson-clover matrix upon each spinor
  involves a mixing of all color and spin components.}  Thus in half
precision a spinor is stored as 6 {\em short4} arrays and a single {\em float}
normalization array.

\subsection{Mixed-precision Linear Solvers}
The use of mixed-precision iterative refinement for solving linear
equations is fairly commonplace on GPUs and other architectures where
the use of double precision comes with a significant performance
penalty.  Such an approach allows the bulk of the computation to be
performed in fast low precision, with periodic updates in high
precision to ensure accuracy of the final solution. Even on
architectures where there is parity between peak single and double
precision performance, a factor of two difference in memory traffic is
unavoidable, and so for bandwidth-bound problems such as our sparse
matrix-vector product, the use of mixed precision remains advantageous.
QUDA uses a variant of reliable updates~\cite{Sleijpen:1996} to
implement mixed-precision iterative refinement.  This approach
has the advantage that a single Krylov space is preserved throughout the
solve, as opposed to the traditional approach of defect correction
which explicitly restarts the Krylov space with every correction,
increasing the total number of solver iterations~\cite{Clark:2009wm}.
It was found that the best time to solution is typically obtained
using either double-half or single-half approaches.

\subsection{Auto-tuned Linear Algebra Kernels}
QUDA provides the additional vector-vector linear algebra (BLAS1-like)
kernels needed to implement the linear solvers.  These additional
routines take advantage of kernel fusion wherever possible to reduce
memory traffic and hence improve performance of the complete solver.
Since each of these kernels and their various half, single, and double
precision variants may have different optimal CUDA
parameters (i.e., sizes of the thread blocks and the number of blocks
treated at once), an auto-tuning approach is taken to ensure maximum
performance.  All possible combinations of parameters are tested for
each kernel, and the optimal values are written out to a header file
for inclusion in production code after a recompilation of the library.
Due to the memory bandwidth intensity of these (essentially streaming)
kernels, the complete solver typically runs 10 to 20\% slower than
would the matrix-vector product in isolation.

\section{Multi-GPU Implementation}
\label{sec:multi-gpu}

\subsection{General Strategy} 

In parallelizing across multiple GPUs, we have taken
the simplest approach by only dividing the time dimension, with
the full extent of the spatial dimensions confined to a single GPU.
This approach was motivated by the asymmetric nature of the lattice
dimensions under study (\(24^3\times128\) and \(32^3\times256\)), and
in order to simplify this initial parallelization.  In this form,
since we are parallelizing over the slowest running spacetime
index, the changes required to the single GPU kernel code were relatively
minimal.  If one were to attempt to scale to hundreds of GPUs or
more, multi-dimensional parallelization would clearly be needed to keep
the local surface to volume ratio under control.  Given current lattice
sizes, however, such extreme parallelization would imply small local
volumes and require rethinking of the fundamental algorithms.  Work
in this direction is underway.

For parallelizing across multiple GPUs, each GPU can either be
controlled using a distinct CPU thread or with a distinct process.
The potential advantage of the threaded approach is that it avoids
unnecessary copies within a node; however, this advantage has
decreased on recent CPUs that feature integrated memory controllers and
much higher memory bandwidth (compared to pre-Nehalem Xeons, for example),
reducing the overhead of an additional local memory
copy.  To communicate between GPUs on different nodes, a message
passing approach is necessary since the memory space is by definition
separate.  While mixed-mode programming is possible (threads within a
node, message passing between the nodes), we exclusively used a
message passing approach since initial investigations suggested no
improvement would be gained from the use of threads.  In particular, we
used QMP (QCD Message Passing) which is an API built on top of MPI
that provides convenient functionality for LQCD computations~\cite{qmp}.

In parallelizing the action of the Wilson-clover matrix onto a spinor
field partitioned between \(N\) distinct GPUs, we slice the temporal
dimension into \(N\) equal sized volumes of size \(V/N = V_sT/N\).
Referring to (\ref{eq:M}), the only part of the matrix that connects
different lattice sites is the action of \(D_{x,x'}\), since the
clover matrix \(A_x\) is local to a given lattice site.  When updating
the sites on either end of the local temporal boundary, the adjacent
spinors which are on the neighboring GPUs are required, as well as the
gauge field matrix connecting these sites.

\subsection{Gauge Field Ghost Zone}
The link matrix \(U_x^\mu\) connecting sites \(x\) and \(x+\hat{\mu}\)
is stored at site \(x\); hence the required link matrix for the
receive from the forward temporal direction for sites in the last
spatial volume (or timeslice) will already be present locally, and
only the adjacent spinor is required.  For the receive from the
backward temporal direction into the first timeslice, the required
link matrix will be on the adjoining GPU and so must be transferred.
Since the link matrices are constant throughout the execution of the
linear solver, we transfer the adjoining link matrices in the program
initialization.  Compared to the original single GPU code, this posed
the obvious question: where should the extra face (the ghost zone) of
gauge field matrices be stored?  Given that the fields were already
padded by an extra spatial volume, a very natural location is within
the padded region since this is exactly the correct size to store the
additional gauge field slice (see Fig.~\ref{fig:layout}).  Altering
the kernel for this change simply required that if the thread id
corresponded to the first timeslice (local to the GPU) then the gauge
field array indices are set to the padded region.  Extra constants
were introduced to describe the boundary conditions at the start and
end of the local volume, since one of these boundaries might
correspond to a global boundary and not just a local boundary.

\subsection{Spinor Field Ghost Zone}

\begin{figure}[ht]
\begin{center}
\includegraphics[width=3in]{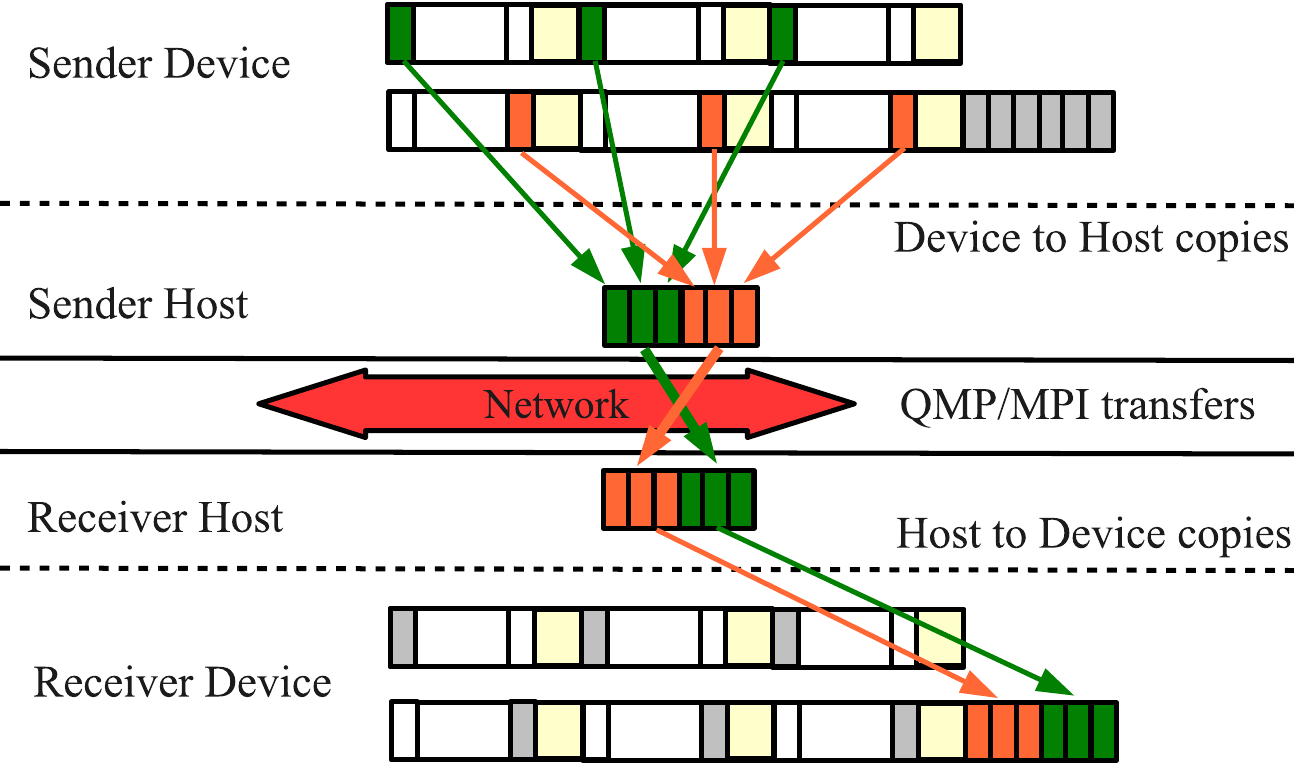}
\end{center}
\caption{\label{fig:comms}Spinor ghost zones and communication steps:
  We show the source spinor on the sending device (top) assuming
  $N_{vec}=4$, corresponding to 6 blocks from
  Fig.~\ref{fig:layout}. The grey buffers at the end correspond to the
  ghost zones. The top 3 blocks correspond to the $P^{+4}$ projected
  components, while the lower 3 blocks nearer the ghost zone
  correspond to $P^{-4}$. Data from the back faces (green) needs to be
  gathered into a communications buffer on the sending host and
  likewise for the forward face (orange). The faces are then
  transferred to the receiving host via QMP/MPI. Once transferred the
  faces are transferred to the ghost zones on the receiving device
  (bottom of diagram), which then uses the data directly from the
  ghost zones, hence the corresponding faces have been greyed out.}
\end{figure}

Our initial strategy for storing the transferred faces was to put them
in the padded regions of the destination GPU's spinor field.  Like the
gauge field ghost zone this seems very natural, but it introduced
complications into the reduction kernels used in the Krylov solvers:
these assume a contiguous memory buffer, and so without careful
rewriting the ghost zones would be double counted.  The approach we
opted for instead was to oversize the spinor fields by the size of the
two transferred faces.  When doing reductions, this end zone can be
simply excluded ensuring correctness.  As described in
Section~\ref{sec:similar} the spin projectors in the temporal direction are
diagonalized, halving the amount of data that needs to be transferred
in the temporal gathering,\footnote{We note that it is true in general
  (for all directions) that only 12 numbers need be transferred,
  regardless of whether or not the projector has been diagonalized.
  In the general case one has to apply the spin projector explicitly
  to the 24 components to obtain 12 components before initiating the
  transfer.  In this special case, since the spin projector is
  diagonal, we merely need to copy the first (second) 12 components
  directly for the positive (negative) projector.} and so the extra
total storage required is actually only \(24V_s\) components. The
upper 12 spinor components which arise from the receive from the
backward direction occupy the first half of the end zone, and the
lower 12 spinor components arising from the receive from the forward
direction occupy the second half.  For half precision the extra
normalization constant for each (12 component) spinor is also
required, and hence an end zone of size \(2V_s\) elements is added to
the normalization field. We illustrate the spinor ghost zones and the
basic communication requirements in Fig.~\ref{fig:comms}.

With the ghost zone elements stored in the end zone, extra indexing
logic was required to ensure that the correct spinors would be loaded by
the threads updating the boundaries.  Fortunately, this extra logic
introduced minimal overhead since warp divergence is avoided
because the number of spatial sites \(V_s\) is divisible by the warp
size, a condition that is met by the lattice dimensions under
consideration here (and all production LQCD calculations that we aware
of).

\subsection{Communication Strategies}

\subsubsection{No Overlap}

The first and simplest approach to parallelization is to perform all
of the communications up front and then do the computation for the
entire volume in a single kernel. The device-to-host
transfers are achieved through the use of separate {\it cudaMemcpy}
calls (one for each face block), with half precision requiring an
extra {\it cudaMemcpy} for the face of the normalization array.  Once
on the host, all of these blocks are contiguous in memory, allowing
for a single message passing in each direction.  The received faces
are sent to the device using a single {\it cudaMemcpy} for each face (with
an extra {\it cudaMemcpy} required for each of the normalization faces in
half precision) and placed in the end zone of the spinor field.
Finally the Wilson-clover kernel is executed.

\subsubsection{Overlapping Communication and Computation}

Our second implementation aimed to overlap all of the communication
with the computation of the internal volume.  To do so, the CUDA
streaming API was used, which allows for a CUDA kernel to execute
asynchronously on the GPU at the same time that data is being
transferred between the device and host using
{\it cudaMemcpyAsync}.\footnote{The Fermi architecture improves upon this
  model by allowing for bidirectional transfers over the PCI-E bus.}

Additionally this makes use of non-blocking MPI communication
possible: after the backward face has been transferred to the host,
the MPI exchange of this face to its neighbor is overlapped with the
transfer of the forward face from device to host.  In turn, when the
first face has been received, this can be sent to the device while the
second face is being communicated. This approach requires three CUDA
streams: one to execute the kernel on the internal volume, one for the
face send backward / receive forward, and one for the face send
forward / receive backward.  An additional required step is that the
streams responsible for gathering the faces to the host must be
synchronized, using {\it cudaStreamSynchronize}, before message
passing can take place to ensure transfer completion.  In principle,
we could also overlap the host-to-device transfer of the second face
and the computation involving the first face.  This would yield a
minimal speedup at best, since the time spent executing the face
kernel is not the limiting factor, and it may actually reduce overall
performance since the kernel would be updating half as many sites at a
time, reducing parallelism and potentially decreasing kernel
efficiency.

\subsection{Parallelizing the Linear Solver}
Aside from the parallelization of the sparse matrix vector product,
the only other required addition to the code was the insertion of MPI
reductions for each of the linear algebra reduction kernels.

\section{Solver Performance Results}\label{sec:solver-perf}

\subsection{Details of the Numerical Experiments}
Our numerical experiments were carried out on the ``9g'' cluster at
Jefferson Laboratory. This cluster is made up of 40 nodes containing 4
GPUs each, as well as an additional 16 nodes containing 2 GPU devices
each that are interconnected by QDR InfiniBand on a single switch. In
this study, we focused our attention primarily on the partition made
up of the 16 InfiniBand connected nodes, with one or two exceptions.
The nodes themselves utilize the Supermicro X8DTG-QF motherboard
populated with two Intel Xeon E5530 (Nehalem) quad-core processors
running at 2.4 GHz, 48 GiB of main memory, and two NVIDIA GeForce GTX
285 cards with 2 GiB of device memory each.

The nodes run the CentOS 5.4 distribution of Linux with version 190.29
of the NVIDIA driver.  The QUDA library was compiled with CUDA 2.3 and
linked into the Chroma software system using the Red Hat version
4.1.2-44 of the GCC/G++ toolchain.  Communications were performed
using version 2.3.2 of the QCD Message Passing library (QMP) built
over OpenMPI 1.3.2. In all our tests we ran in a mode with one MPI
process bound to each GPU.
 
The numerical measurements were taken from running the Chroma
propagator code and performing 6 linear solves for each test (one for
each of the 3 color components of the upper 2 spin components), with
the quoted performance results given by averages over these solves.
Statistical errors were also determined but are generally too small to
be seen clearly in the figures.  Importantly, all performance results
are quoted in terms of ``effective Gflops'' that may be compared with
implementations on traditional architectures.  In particular, the
operation count does not include the extra work done to reconstruct
the third row of the link matrix.

We carried out both strong and weak scaling measurements. The strong
scaling measurements used lattice sizes of $V=24^3 \times 128$ and
$V=32^3 \times 256$ sites respectively. Both the lattice sizes and the
Wilson-clover matrix had their parameters chosen so as to correspond
to those in current use by the {\em Anisotropic Clover} analysis
program of the Hadron Spectrum collaboration. The lattices used were
{\em weak field} configurations.  Such configurations are made by
starting with all link matrices set to the identity, mixing in a small
amount of random noise, and re-unitarizing the links to bring the
links back to the $SU(3)$ manifold.  We emphasize that while these
lattices were not physical, we have tested the code on actual
production lattices on both the volumes mentioned for correctness. The
concrete physical parameters do not affect the rate at which the code
executes but control only the number of iterations to convergence in
the solver. The weak scaling tests utilized local lattice sizes of
$V=32^4$ and $V=24^3\times 32$ sites per GPU, respectively.

The solver we employed was the reliably updated BiCGstab solver
discussed in \cite{Clark:2009wm}. We ran the solver in single
precision and mixed single-half precision with and without overlapped
communications in the linear operator. For the lattices with
$V_s=24^3$ spatial sites, we also ran the solver in uniform double
precision and in mixed double-half precision modes. When run in single
or single-half mixed precision modes the target residuum was
$||r||=10^{-7}$, whereas in the double precision and mixed double-half
precision modes the residuum was $||r||=10^{-14}$.  In addition, the
delta parameter was set to $\delta=10^{-3}$ in single,
$\delta=10^{-1}$ in mixed single-half, $\delta=10^{-5}$ in double and
$\delta=10^{-2}$ in the mixed double-half modes of the solver
respectively.  The meanings of these parameters are explained fully in
\cite{Clark:2009wm}.

\subsection{Weak Scaling}

\begin{figure}[htb] 
\subfloat[][]{\includegraphics[width=3.5in]{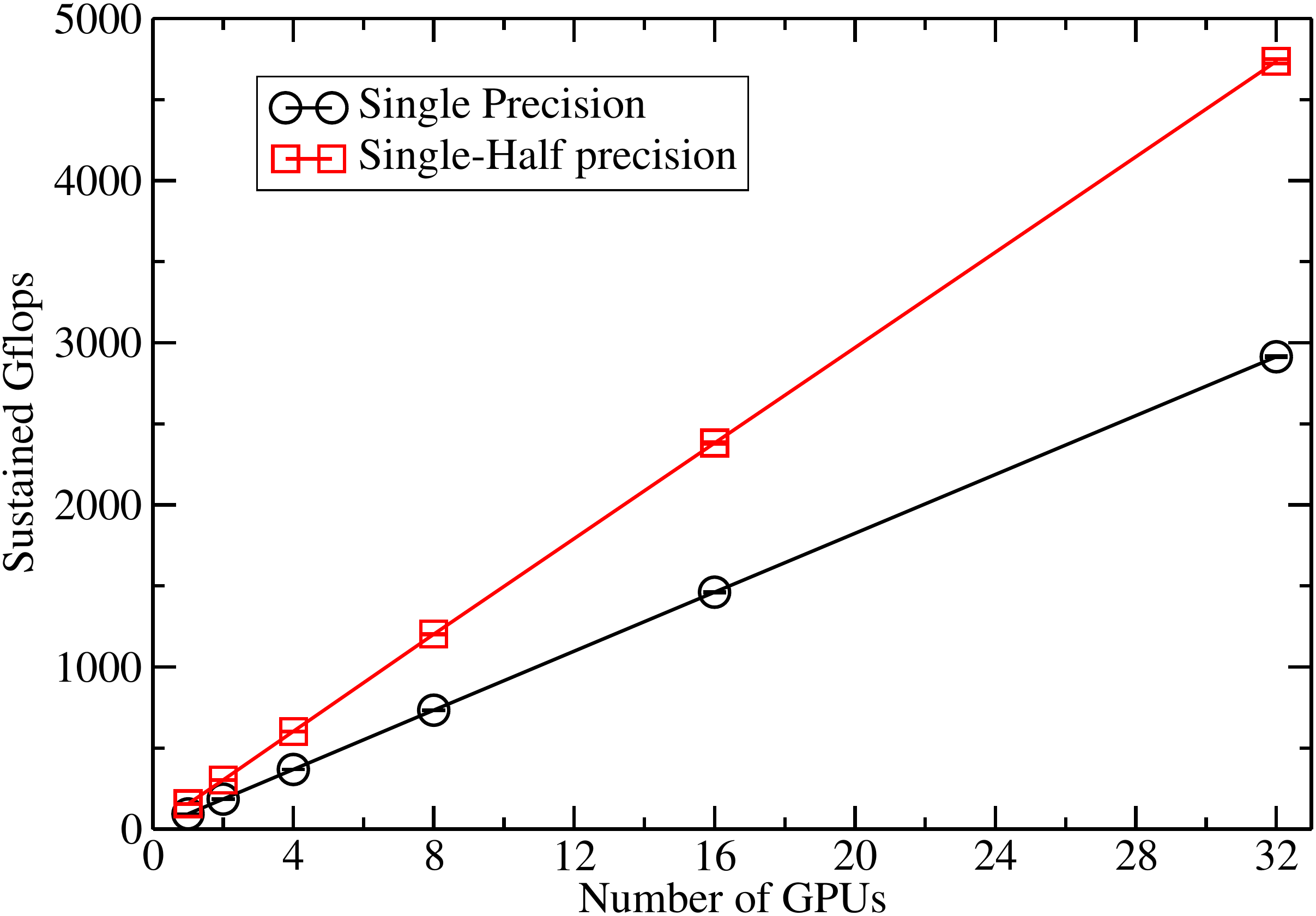}}\\
\subfloat[][]{\includegraphics[width=3.5in]{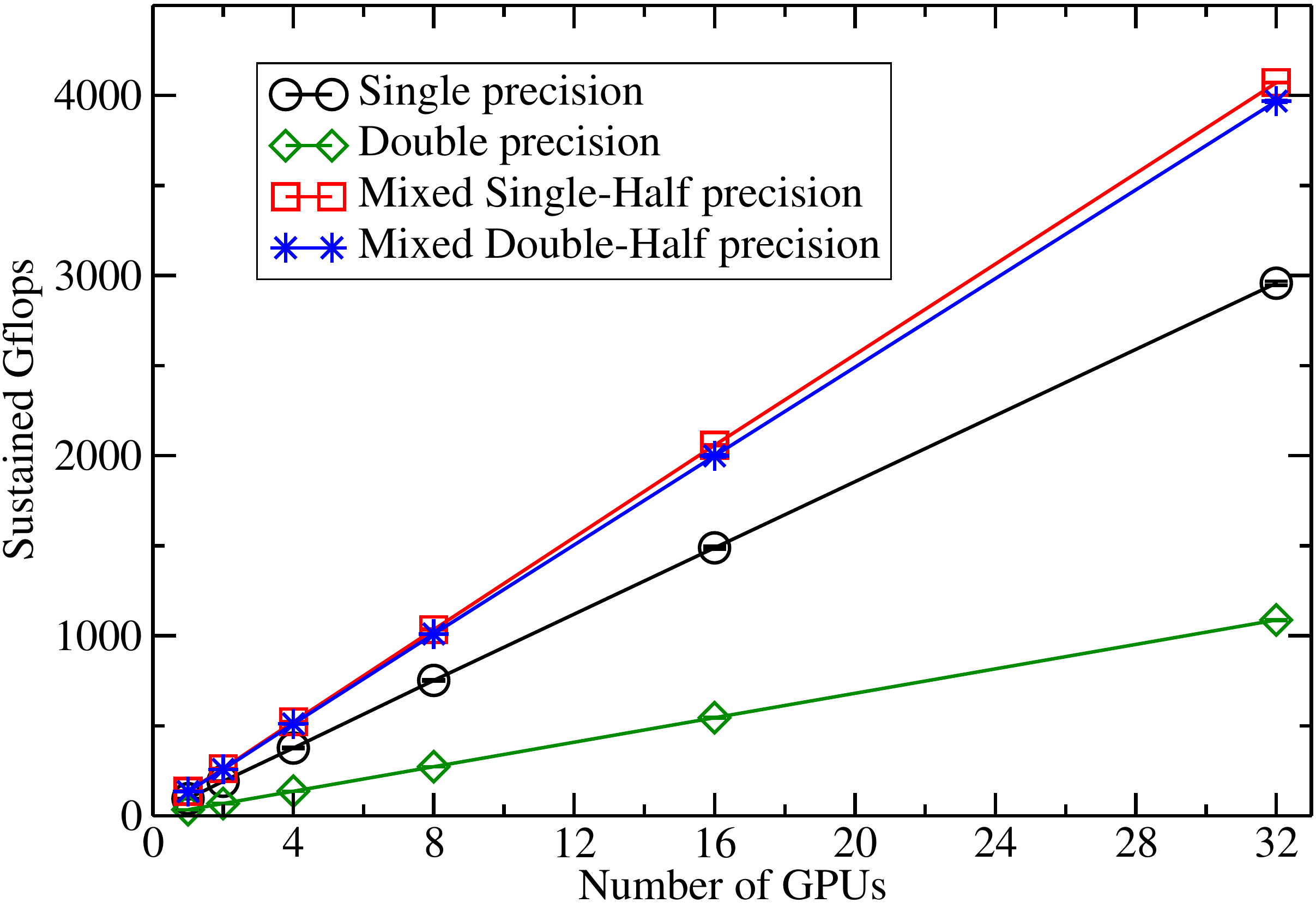}}
\caption{\label{fig:weak-scale} Weak scaling results for up to 32 GPUs
  on lattices with local volumes of (a) $V=32^4$ and (b) $V=24^3
  \times 32$ sites per GPU.  In subfigure (a) we show performance
  results for the single precision solver and the mixed single-half
  precision solver.  In subfigure (b) we also show results for double
  precision and mixed double-half precision. In both (a) and (b), the
  data come from solvers where communications and computation have
  been overlapped, as this performed fastest in weak scaling tests.}
\end{figure}

Our results for weak scaling are shown in
Fig.~\ref{fig:weak-scale}. We see near linear scaling on up to 32 GPUs
in all solver modes.  In the case with $V=32^4$ sites per GPU, we were
unable to fit the double precision and mixed double-half precision
problems into device memory, and hence we show only the single and
single-half data. In the case with local volume of $24^3\times32$ we
show also double precision and mixed double-half precision data. It is
gratifying to note that the mixed double-half precision performance of
Fig.~\ref{fig:weak-scale}(b) is nearly identical to that of the
single-half precision case. Both mixed precision solvers are
substantially more performant than either the uniform single or the
uniform double precision solver.  We note that for lattices with $32^4$
sites per GPU we have reached a performance of 4.75 Tflops.

\subsection{Strong Scaling}

\begin{figure}[htb] 
\subfloat[][]{\includegraphics[width=3.5in]{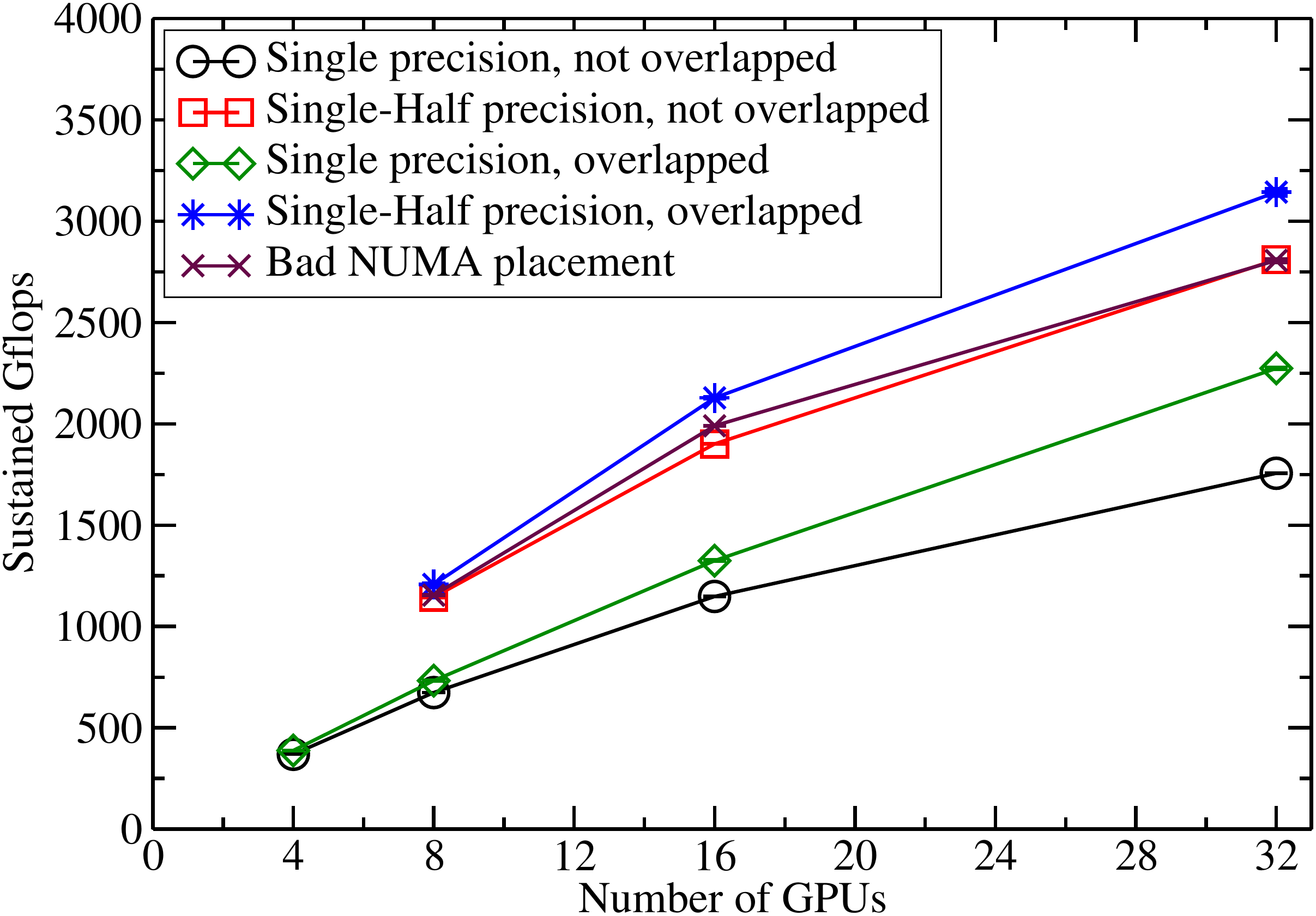}} \\
\subfloat[][]{\includegraphics[width=3.5in]{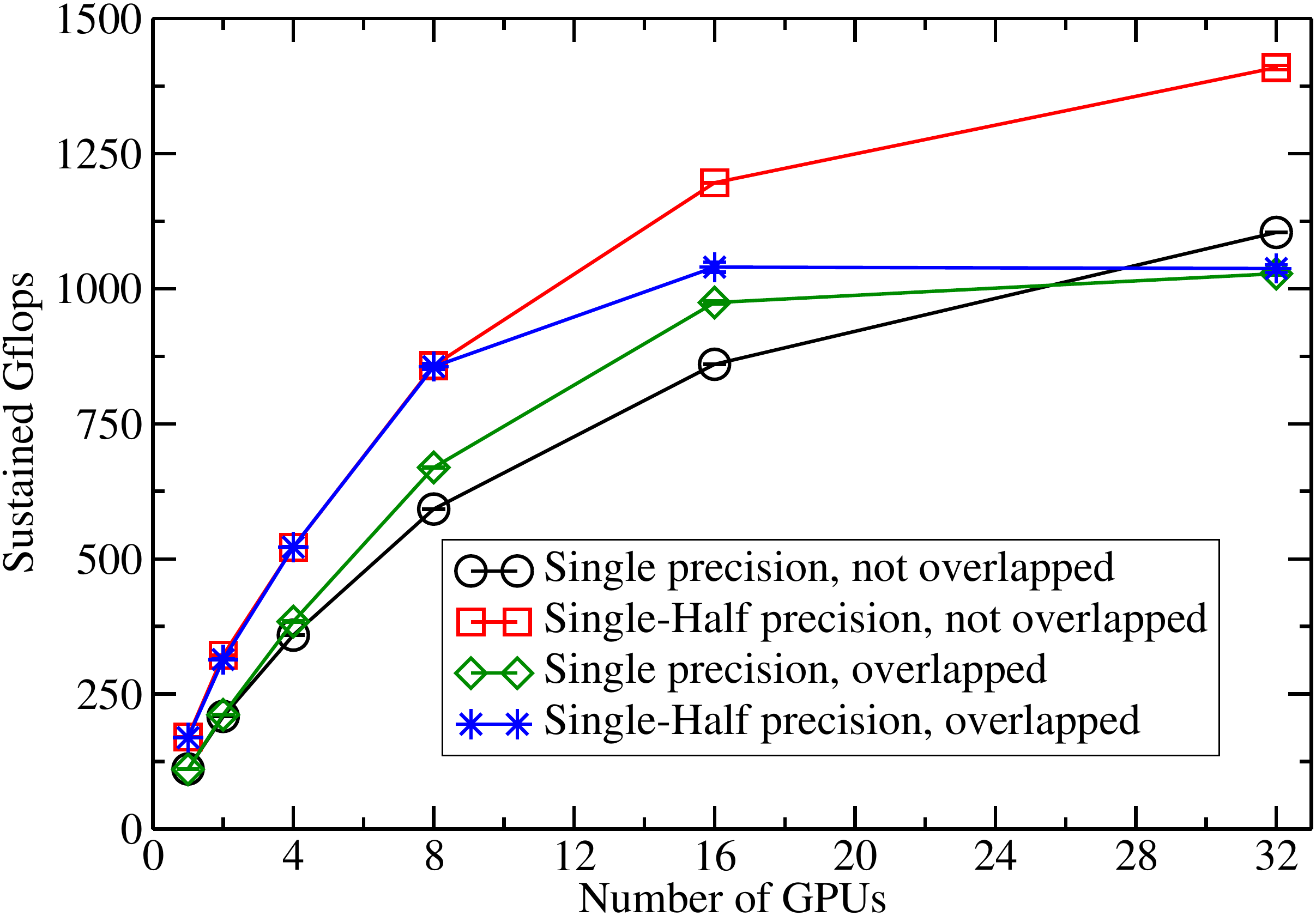}} \\
\caption{\label{fig:strong-scale} Strong scaling results for up to 32
  GPUs on lattices with total volumes of (a) $V=32^3 \times 256$ sites
  and (b) $V=24^3 \times 128$ sites.  We show data for single precision
  without overlapping communications and computation (black), mixed
  single-half precision without overlapping communications and
  computation (red), single precision with overlapped communications
  and computation (green), and mixed single-half precision with overlapped
  communications and computation (blue).  In subfigure (a) we also show
  mixed single-half precision with overlapped communications but
  with deliberately bad NUMA placement (maroon).}
\end{figure}

\begin{figure}[htb]
\includegraphics[width=3.5in]{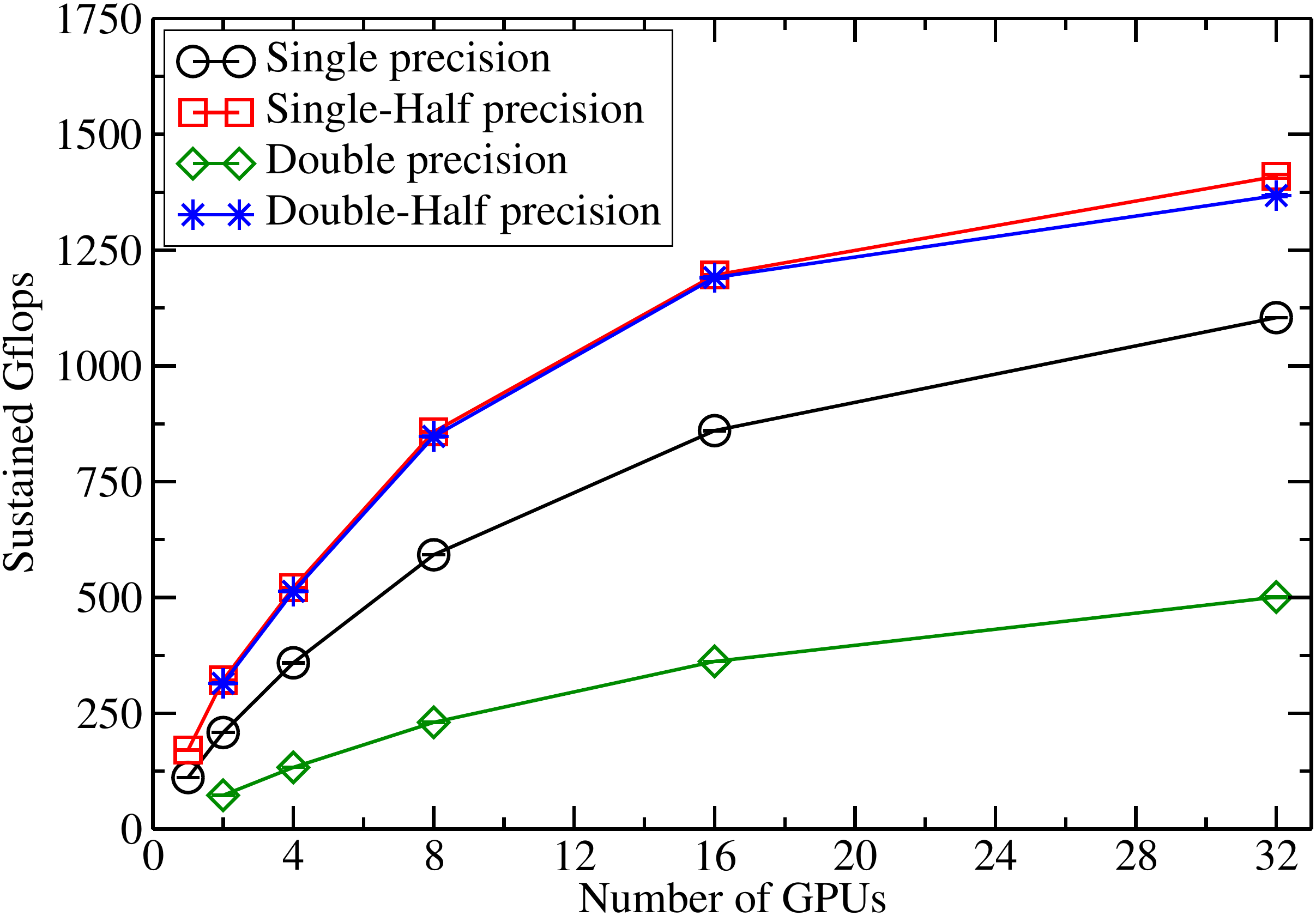}
\caption{\label{fig:strong-scale24}Strong scaling results for the
  $V=24^3\times 128$ lattice in single precision, double precision,
  single-half mixed precision, and double-half mixed precision.  We
  used the solver that did not overlap computations and communications
  for these results, since as shown in Fig.~\ref{fig:strong-scale} it
  was faster than the overlapped solver for the $V=24^3\times 128$
  lattice in single and mixed single-half precisions.}
\end{figure}

Fig.~\ref{fig:strong-scale} shows our strong scaling results.  In
Fig.~\ref{fig:strong-scale}(a) we show the data for the lattices with
$V=32^3\times 256$ sites. We see a clear deviation from linear scaling
as the number of GPUs is increased and the local problem size per GPU
is reduced. This is not unexpected, since as the number of GPUs is
increased the faces represent a larger fraction of the overall work.
The improvement from overlapping communication with computation is
increasingly apparent as the number of GPUs increases.  The benefits
of mixed precision over uniform single precision can clearly be seen.
However, we note that performing the mixed precision computation comes
with a penalty in terms of memory usage: the mixed precision solver
must store data for both the single and half precision solves, and
this increase in memory footprint means that at least 8 GPUs are
needed to solve this system.  The uniform single precision solver
requires only the single precision data and can be solved (at a
performance cost) already on 4 GPUs. We highlight the fact that the 32
GPU system is made up of 16 cluster nodes, which themselves contain
128 Nehalem cores.  We have performed a solution of this system on the
Jefferson Lab ``9q'' cluster, which is identical in terms of cores and
InfiniBand networking but does not contain GPUs.  On a 16-node
partition of the ``9q'' cluster we obtained 255 Gflops in single
precision using highly optimized SSE routines, which corresponds to
approximately 2 Gflops per CPU core. In our parallel GPU computation,
on 16 nodes and 32 GPUs we sustained over 3 Tflops which is over a
factor of 10 faster than observed without the GPUs.

Fig.~\ref{fig:strong-scale}(b) shows our strong scaling results for
the lattice with $V=24^3 \times 128$ sites. This lattice has half the
time extent of the larger lattice, and thus we expect strong scaling
effects to be noticeable at smaller GPU partitions than in the
previous case. Further, the spatial volume is a factor of $2.3$
smaller for the $V=24^3 \times 128$ lattices than for the larger
case. We were surprised that the trend in our results is different
from that in Fig.~\ref{fig:strong-scale}(a). Notably, in this case we
seem to gain little from overlapping communication and computation in
the mixed precision solver. Indeed, for more than 8 GPUs the mixed
precision performance reaches a plateau and is surpassed even by the
purely single precision case. We believe this dropoff in the strong
scaling is due to additional overheads incurred in overlapping
communications with computations arising from system and driver
issues. We will return to this point in Section~\ref{sec:system},
where we discuss latency microbenchmarks, but suffice it to say that
using {\em cudaMemcpyAsync} appears to have a higher latency than {\em
  cudaMemcpy}. This may be a feature of our motherboard or the version
of the NVIDIA driver we are using. In the case of the $V=32^3\times
256$ lattice, probably the volume in the body is large enough to hide
this extra latency. In the case of the $V=24^3\times 128$ lattice, our
data suggests that the local volume may be sufficiently small that the
overhead of setting up the asynchronous transfers dominates and that
in this instance the lower latency of synchronous {\em cudaMemcpy}
calls can result in better performance.

Fig.~\ref{fig:strong-scale24} shows the strong scaling data for
various precision combinations for the $V=24^3\times 128$ lattice,
where we now include uniform double and mixed double-half precision
results and do not overlap communication with computation. Again we
see that the mixed precision solvers employing half precision
outperform both single and double uniform precision solvers.  Note
that uniform double precision exhibits the best strong scaling of all
because this kernel is less bandwidth bound due to the much lower
double precision peak performance of the GTX 285 (see Table
\ref{table:specs}).

\subsection{System Issues}\label{sec:system}

The PCI-E architecture in our Supermicro nodes was such that the two
GPU devices were each on a bus with a direct connection to a separate
socket on the motherboard. In our tests we launched two MPI processes
per node. In order to obtain maximum bandwidth on the buses, it was
necessary to explicitly bind each MPI process to the correct socket.
We accomplished this using the processor affinity feature of OpenMPI.

In Fig.~\ref{fig:strong-scale}(a) we show the performance a
deliberately badly chosen NUMA configuration (with maroon
x-symbols). We bound each process to the opposite socket from the CUDA
device it was using. One can see that the performance is noticeably
lower than the correctly bound case denoted by blue asterisks in
Fig.~\ref{fig:strong-scale}(a).

Secondly, as alluded to previously, we note that on these nodes the
latencies of {\em cudaMemcpy} (used in the non-overlapped
communication code) and of {\em cudaMemcpyAsync} (followed immediately
by a {\em cudaSynchronizeThread}) call are quite different.

\begin{figure}[htb]
\includegraphics[width=3.5in]{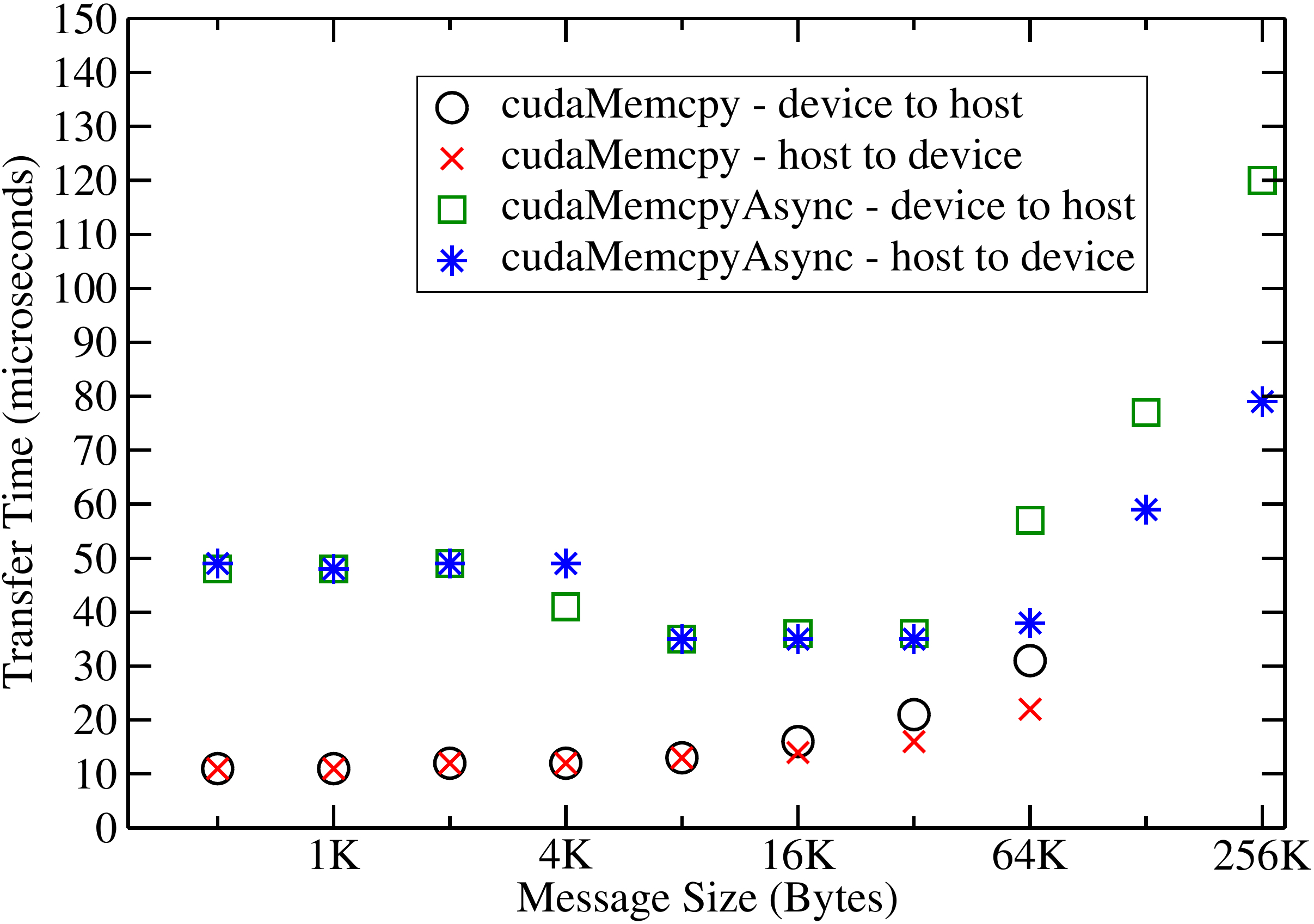}
\caption{\label{fig:latency}Latency microbenchmark showing tranfer times from host to device or vice versa for messages of varying sizes. We show data for: device to host using {\em cudaMemcpy} (black), host to device using {\em cudaMemcpy} (red), device to host using {\em cudaMemcpyAsync} + {\em cudaSynchronizeThreads} (green) and host to device using {\em cudaMemcpyAsync}+{\em cudaSynchronizeThreads} (blue). The timings are taken over 500,000 message transfers }
\end{figure}

As shown in Fig.~\ref{fig:latency}, using {\em cudaMemcpyAsync} incurs
a latency of just under 50 microseconds whereas a synchronous {\em
  cudaMemcpy} has a much shorter latency of 11 microseconds.  It can
also be seen that once out of the latency limited region, the graphs
show different gradients for the host-to-device and device-to-host
transfers, indicating different host-to-device and device-to-host
bandwidths.  These features may depend somewhat on the version of the
NVIDIA driver and motherboard BIOS used, but additional testing so far
suggests that the main culprit is a hardware limitation in the early
revision of the Intel 5520 (Tylersburg) chipset used in the nodes.
Therefore the decision on whether to overlap communication and
computation or not may depend on the system under consideration, as
well as the problem size.

\section{Conclusion and Future Work}\label{sec:conclusions}

We have demonstrated what we believe is the first successful attempt
to use multiple GPU units in parallel for LQCD computations. We have
weak scaled our application to 4.75 Tflops on 32 GPUs and have strong
scaled the application, on a problem size of scientific interest, to
over 3 TFlops. In this latter case, we have achieved over a factor of
10 increase in performance compared to not using GPUs (255 Gflops on a
``regular'' cluster partition containing the same number processors).
We believe that the order of magnitude increase in computing power is
an enabling technology for sophisticated modern analysis methods of
great interest to particle and nuclear physics. Indeed the solver we
have described is now in use in production LQCD calculations of the
spectrum of hadrons using the technique of {\em distillation}
\cite{Peardon:2009gh,Dudek:2010wm}. Current calculations use lattice
configurations of the same size as described in
Section~\ref{sec:solver-perf} which were generated on leadership
computing platforms under DOE INCITE and NSF TeraGrid allocations
(granted to the USQCD and Hadron Spectrum collaborations,
respectively). The calculations involve 32768 calls to the solver for
each configuration and benefit enormously from the speedup delivered by
the GPU solver.

Prior to parallelizing the QUDA library, our larger volume dataset was
not amenable to solution on GPUs due to memory constraints.
The use of multiple GPUs allows the solution to proceed, realizing
the large increases in cost effectiveness promised by GPUs.

A slightly more nuanced point is that the nodes containing 4 GPUs (and
no InfiniBand) may now be more efficiently utilized. Prior to
parallelization, one could solve the $24^3 \times 128$ problem on a
single GPU and analyze two configurations simultaneously on a single
node. One could not analyze more, due to the limitations on the host
(primarily memory capacity). Now one can analyze 2 configurations
simultaneously using 2 GPUs each, optimally utilizing all 4 GPUs in
the node. The exact optimization of a node configuration in terms of
InfiniBand cards, GPUs, and operating model is an interesting issue but
is beyond the scope of this paper.

There are many avenues for future exploration. Currently only the
solvers have been accelerated in the QUDA library. Parallelization
onto multiple GPUs may make gauge generation on GPU clusters an
interesting and desirable possibility. We are also interested in
porting more modern algorithms to the GPUs such as the adaptive
multigrid solver discussed in \cite{Brannick:2007ue} to speed up
computations even further. We follow the development of the OpenCL
standard with interest with a view to potentially harness GPU devices
from AMD as well as NVIDIA, and we await future hardware and software
improvements to allow better coexistence of GPUs and message-passing
(such as sharing pinned memory regions between CUDA and MPI).  Finally,
we hope that the lessons learned from GPUs will be usefully applicable
on heterogeneous systems in general as we head towards the exascale.

\section*{Acknowledgment}

The authors would like to thank Chip Watson for funding an extremely
productive week of coding, and for dedicated access to the Jefferson
Lab 9g cluster.  Enlightening discussions with Jie Chen, Paulius
Micikevicius, and Guochun Shi are also gratefully acknowledged.  This
work was supported in part by U.S.\ NSF grants PHY-0835713 and
OCI-0946441 and U.S.\ DOE grant DE-FC02-06ER41440.  Computations were
carried out on facilities of the USQCD Collaboration at Jefferson
Laboratory, which are funded by the Office of Science of the
U.S.\ Department of Energy.  Authored by Jefferson Science Associates,
LLC under U.S.\ DOE Contract No.\ DE-AC05-06OR23177.  The
U.S.\ Government retains a non-exclusive, paid-up, irrevocable,
world-wide license to publish or reproduce this manuscript for
U.S.\ Government purposes.

\IEEEtriggeratref{12}


%

\end{document}